\documentclass[pre,aps,twocolumn]{revtex4}

\usepackage{graphicx}
\usepackage{amsmath}
\usepackage{amssymb}
\usepackage{ifthen}
\usepackage{booktabs}
\usepackage{gensymb}
\usepackage{color}

\usepackage{graphicx}
\usepackage{dcolumn}
\usepackage{bm}
\usepackage{longtable}

\newcommand{\bb}{\begin{equation}}
\newcommand{\ee}{\end{equation}}
\newcommand{\ba}{\begin{eqnarray*}}
\newcommand{\ea}{\end{eqnarray*}}
\newcommand{\rhor}{\rho({\bf r})}

\newcommand{\rr}{{\mathbf r}}
\newcommand{\dr}{{\rm d}{\bf r}}

\begin{document}

\title{Edge contact angle and modified Kelvin equation for condensation in open pores}

\author{Alexandr \surname{Malijevsk\'y}}
\affiliation{ {Department of Physical Chemistry, Institute of
Chemical Technology, Prague, 166 28 Praha 6, Czech Republic;}\\
 {Laboratory of Aerosols Chemistry and Physics, Institute of Chemical Process Fundamentals, Academy of Sciences, 16502 Prague 6, Czech Republic}}
\author{Andrew O. \surname{Parry}}
\affiliation{Department of Mathematics, Imperial College London, London SW7 2BZ, UK}
\author{Martin \surname{Posp\'\i\v sil}}
\affiliation{Department of Physical Chemistry, Institute of Chemical Technology, Prague, 166 28 Praha 6, Czech Republic}

\begin{abstract}

We consider capillary condensation transitions occurring in open slits of width $L$ and finite height $H$  immersed in a reservoir of vapour. In this
case the pressure at which condensation occurs is closer to saturation compared to that occurring in an infinite slit ($H=\infty$) due to the
presence of two menisci which are pinned near the open ends. Using macroscopic arguments we derive a modified Kelvin equation for the pressure,
$p_{cc}(L;H)$, at which condensation occurs and show that the two menisci are characterised by an edge contact angle $\theta_e$ which is always
larger than the equilibrium contact angle $\theta$, only equal to it in the limit of macroscopic $H$. For walls which are completely wet ($\theta=0$)
the edge contact angle depends only on the aspect ratio of the capillary and is well described by $\theta_e\approx \sqrt{\pi L/2H}$ for large $H$.
Similar results apply for condensation in cylindrical pores of finite length. We have tested these predictions against numerical results obtained
using a microscopic density functional model where the presence of an edge contact angle characterising the shape of the menisci is clearly visible
from the density profiles. Below the wetting temperature $T_w$ we find very good agreement for slit pores of widths of just a few tens of molecular
diameters while above $T_w$ the modified Kelvin equation only becomes accurate for much larger systems.
\end{abstract}

\maketitle

It is well known that within a narrow capillary slit of width $L$ the forces of surface tension shift the phase boundary for coexistence between
vapour and liquid $p_{cc}$ away from saturation pressure $p_{\rm sat}$. If both the height and depth of the two facing walls forming the slit are
infinitely large the phase boundary shift $\delta p_{cc}=p_{\rm sat}-p_{cc}$ at which capillary coexistence occurs is very accurately described by
the macroscopic Kelvin equation \cite{thomson}
 \bb
 \delta p_{cc}(L)\approx\frac{2\gamma\cos\theta}{L}\,. \label{kelvin}
 \ee
Here, $\gamma$ is the liquid-gas surface tension while $\theta$ is the equilibrium contact angle as given by Young's equation,
$\gamma_{wg}=\gamma_{wl}+\gamma\cos\theta$, where $\gamma_{wl}$ and $\gamma_{wg}$ are the wall-liquid and wall-gas surface tensions, respectively.
Microscopic studies of capillary condensation based on classical density functional theory (DFT) \cite{evans79, tarazona83, evans86} and computer
simulation \cite{gelb99} have shown that the Kelvin equation is remarkably accurate, particularly for partial wetting even down to slit widths of
molecular dimensions.  More recently, it has been realized that capping a capillary at one end, equivalent to the construction of a deep groove, may
significantly alter the order of the capillary condensation \cite{rascon07, roth11,mal12, rascon13, petr13, mal14}. This is because capping a
capillary necessitates the formation of a single meniscus which unbinds from the capped end either continuously or discontinuously as the pressure is
increased towards capillary condensation.

\begin{figure}[h]
\includegraphics[width=7cm]{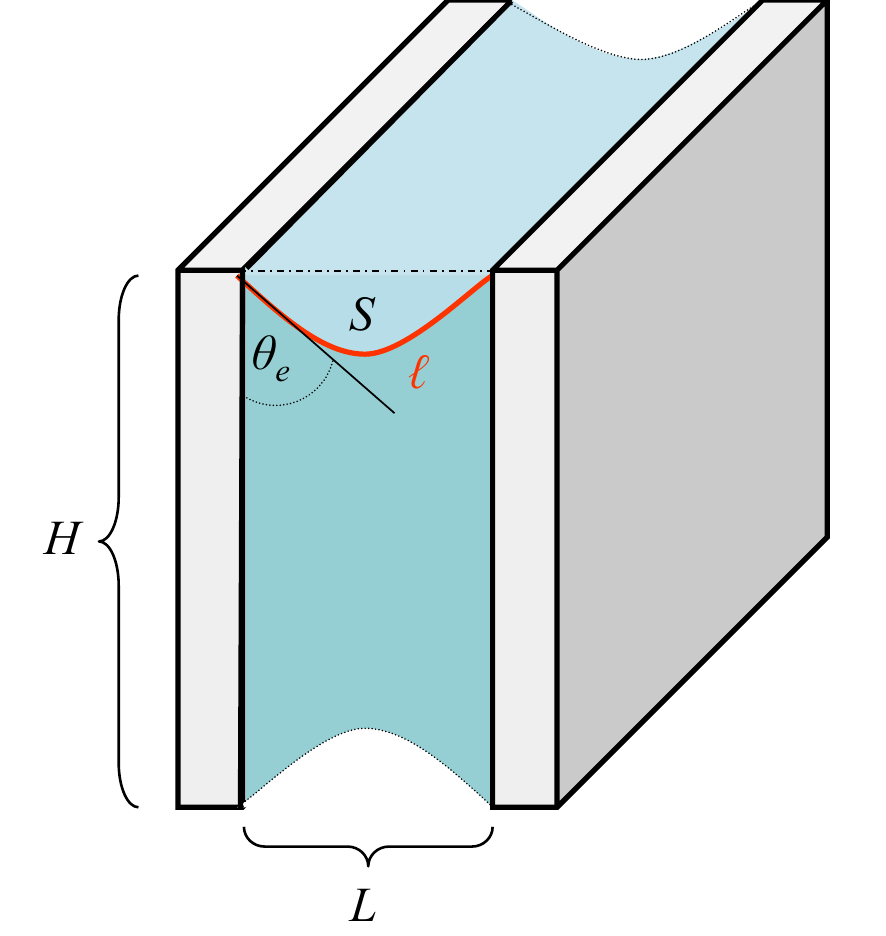}
\caption{Sketch of an open capillary of macroscopic depth but finite width $L$ and finite height $H$.
In the condensed phase two menisci form at the open ends separating the liquid inside the capillary and the reservoir outside.
The menisci are characterised by an edge contact $\theta_e>\theta$ because the pressure at which capillary condensation occurs is
 necessarily closer to saturation compared to that for an infinite slit ($H=\infty$). Here $\ell$ is the meniscus circumference
 while $S$ is the exposed area of vapour (per unit length). }
\end{figure}

In this paper we show that the concept of a  contact angle must be modified when capillary condensation occurs in an open slit of finite height $H$
(see Fig.~1). In this case the condensed liquid-like phase has two menisci, pinned near the open ends, which we show are characterized by an edge
contact angle $\theta_e(H)$ which is {\it{greater}} than the contact angle $\theta$, as defined via Young's equation for a drop on a planar wall. In
particular, for conditions of complete wetting ($\theta=0$) the macroscopic theory predicts that the edge contact angle takes a universal value
determined only by the aspect ratio $L/H$ with $\theta_e\approx\pi/2$ when $L/H$ is of order unity. The shift in the contact angle in turn leads to a
correction to the Kelvin equation of order $1/H$ implying that condensation occurs closer to saturation pressure than for an infinite slit. To test
these predictions we have studied condensation in an open slit using an accurate model density functional (DFT) based on Fundamental Measure Theory
\cite{rosenfeld}. Our numerical results for the pressure at capillary condensation as well as the value of the edge contact angle, for conditions of
both partial and complete wetting are in good agreement with analytic predictions of the macroscopic theory. Predictions for condensation in  a
cylindrical pore are also presented.

To begin we recall the approximate thermodynamic argument leading to the macroscopic Kelvin equation (1) in an infinite capillary slit. Condensation
occurs when the Grand potentials $\Omega$ (per unit area of one wall, say) of the gas-like and liquid-like phases are equal. The volume (pressure)
and area (surface tension) contributions to each imply that for wide slits $\Omega_g\approx -pL+2\gamma_{wg}$ and $\Omega_l\approx-p^*L+2\gamma_{wl}$
where here $p^*=p-\delta p$ is the shifted pressure of the liquid phase which recall would be metastable in the bulk. Equating the potentials implies
that $\delta p_{cc}=2(\gamma_{wg}-\gamma_{wl})/L$ which reduces to (1) on using Young's equation. The Kelvin equation also has a simple geometrical
interpretation since at capillary condensation any phase separation of the gas-like and liquid-like phases is only possible via the formation of a
meniscus, the position of which can be considered arbitrary. Noting that the Laplace equation requires that this meniscus has a circular
cross-section of radius $R=\gamma/\delta p$ the value of $\delta p_{cc}$ then follows from requiring that the meniscus meets each wall at the
equilibrium contact angle $\theta$.

Let us now extend this argument to an open slit of width $L$ and height $H$ and suppose the system
is immersed in a reservoir of vapour at pressure $p<p_{sat}$ (or chemical potential $\mu$). The walls are considered to be infinitely deep and translational invariance is assumed in this direction. Again, we must consider the Grand potentials of the gas-like and liquid-like phases.
For the gas-like phase the contributions are similar to those for the infinite slit coming from the pressure and wall-gas surface tension. Thus, per
unit wall length, we can write $\Omega_g\approx (-pL+2\gamma_{wg})H$ where we have ignored the contribution from the outside walls which are the same
for both the gas-like and liquid-like phases. However, for the liquid-like phase in addition to the (shifted) pressure and wall-liquid surface
tension contributions there is now the additional surface tension cost associated with the area of the two menisci; recall these must be present in
order to separate the capillary liquid from the reservoir of gas outside the slit. This increase in the Grand potential of the liquid-like phase
means that for all finite $H$ the condensation must occur at a pressure which is closer to saturation than that given by the Kelvin equation (1). As
a consequence, the two circular menisci cannot form a stable configuration within the slit itself - if they did they would have to satisfy Young's
equation and this is only possible at the Kelvin pressure (\ref{kelvin}); at the higher pressure required to condense liquid in the finite depth
slit, the two menisci effectively repel until they are pinned at the open ends. Once pinned here, while still of circular cross-section,  they are no
longer required to have a contact angle satisfying Young's equation. This means that in order to balance the Grand potentials we must allow that the
menisci are characterised by an edge contact angle $\theta_e\ne\theta$. Thus for the liquid-like phase we write

 \bb
 \Omega_l\approx  p^*(LH-2S)- 2p S +2\gamma_{wg}H+2\sigma \ell
 \ee
where $\ell=(\pi-2\theta_e)R$ is the length of arc of the pinned meniscus and $S=(\pi/2-\theta_e)R^2-\sin \theta_e RL/2$ is the area between the
meniscus and the capped end (see Fig.1). Equating the Grand potentials determines the value of $\delta p$ at capillary co-existence and hence the
menisci radius $R=\gamma/\delta p$. However elementary geometry also implies that $R=L/(2\cos\theta_e)$ and combining these two relations determines
the location of the capillary condensation in the finite depth capillary as

\bb
 \delta p_{cc}(L;H)=\delta p_{cc}(L)-\frac{\gamma}{H}\left(\sin\theta_e+\sec\theta_e\left(\frac{\pi}{2}-\theta_e\right)\right) \label{mod_kelvin}
\ee
where $\delta p_{cc}(L)$ is the usual Kelvin equation expression pertinent to the infinite ($H=\infty$) slit. The value of the edge contact angle is itself determined as

\bb
 \cos\theta=\cos\theta_e+\frac{L}{2H}\left(\sin\theta_e+\sec\theta_e\left(\frac{\pi}{2}-\theta_e\right)\right)\,. \label{thetae}
\ee
 which, in combination with (\ref{mod_kelvin}) is the central new result of our article. Thus, as remarked above condensation in the open slit
occurs closer to saturation with the correction to the standard Kelvin equation  $\delta p_{cc}(L)-\delta p_{cc}(L;H)\approx 1/H$ for large $H$. This
is consistent with macroscopic results obtained previously for condensation between square plates and blocks although these do not allow for the
circular shape of the meniscus and the modification of the contact angle \cite{Luzar,Netz,Chacko}. Similarly the result (\ref{thetae}) predicts that
the edge contact angle is always larger than the equilibrium contact angle with $\theta_e\to\theta^+$ as $H\to\infty$. For $H/L\gg 1$, the edge
contact angle has the expansion
\begin{equation}
\theta_e=\theta+\left(1+\frac{\pi-\theta}{\sin\theta}\right)\frac{L}{2H}+\cdots
\end{equation}
provided that $\theta>0$. Notice that the amplitude of the correction term diverges as $\theta\to 0$ implying that greatest change to the value of edge contact angle occur for for complete wetting. In this case $\theta_e$ depends only on the aspect ratio $L/H$ and the solution to (\ref{thetae}) has the asymptotic expansion
\bb
  \theta_e\sim\sqrt{\frac{\pi}{2}\frac{L}{H}}\,,
 \ee
as $L/H\to0$. Note that in the opposite limit, $L/H\to 1$, the edge contact angle for complete wetting $\theta_e\to \pi/2$ implying that the meniscus
at condensation is flat and hence that condensation occurs exactly at bulk saturation, $\delta p_{cc}(L;H)=0$. Smaller heights with $H<L$ cannot be
considered since then $\delta p _{cc}(L;H)<0$ implying that the reservoir vapour itself is metastable.

Similar phenomena occur in other geometries. Consider, for example, the condensation of liquid in a cylindrical pore of radius $\mathcal{R}$ and
finite length $H$. For macroscopic $H$ the standard Kelvin equation predicts that condensation occurs when $\delta
p_{cc}(\mathcal{R})=2\gamma\cos\theta/\mathcal{R}$. When $H$ is finite condensation is once again shifted closer to bulk saturation due to the
free-energy cost of the menisci at the open ends. These are now of spherical shape, with Laplace radius $R=2\gamma/\delta p$, and are characterised
by a different edge contact angle $\theta_e $ pertinent to the cylindrical geometry. Repeating the thermodynamic arguments above, now allowing for
the volume $V=\pi R^3(2+\sin^3\theta_e-3\sin\theta_e)/3$ (analogous to $S$) and area $A=2\pi R^2/(1+\sin\theta_e)$ (analogous to $\ell$) of each
menisci, determines that condensation occurs when $\delta p_{cc}(\mathcal{R};H)=2\gamma\cos\theta_e/\mathcal{R}$ where the cylindrical edge contact
is given by
\begin{equation}
\cos\theta_e=\frac{\cos\theta-\frac{2\mathcal{R}}{H}\sqrt{\sin\theta^2+\frac{4\mathcal{R}^2}{H^2}}}{1+\frac{4\mathcal{R}^2}{H^2}}
\end{equation}
For complete wetting this simplifies to $\cos\theta_e=(H^2-4\mathcal{R}^2)/(H^2+4\mathcal{R}^2)$ which behaves as $\theta_e\approx 4\mathcal{R}/H$
for large $H$. This illustrates that the edge contact is geometry dependent, although the shift of the Kelvin equation is still of order $1/H$.

To test these macroscopic predictions we turn to a microscopic DFT model similar to our previous work on condensation in a capped capillary
\cite{mal13}. Within classical DFT the equilibrium density profile is found from minimization of the Grand Potential functional
\begin{equation}
\Omega[\rho]=F[\rho]-\int d{\bf{r}}(\mu-V({\bf{r}}))\rho({\bf{r}})
\end{equation}
where $V({\bf{r}})$ is the external potential arising from the confinement. Here $F[\rho]$ is the intrinsic Helmholtz free-energy functional of the
one-body density $\rhor$ which is  conveniently written as an exact ideal term and an excess part which must be approximated in model calculations.
Following most modern DFT treatments we write
\begin{equation}
F_{ex}[\rho]=F_{hs}[\rho]+\frac{1}{2}\int\int d{\bf{r}}_1d{\bf{r}}_2\phi_{\rm att}(|\rr_1-\rr_2|)
\label{F}
\end{equation}
which, in the spirit of van der Waals, further splits the excess free-energy into repulsive (hard-sphere) and attractive contributions where for the
latter we employ a simple but reliable mean-field approximation. Thus $\phi_{\rm att}(|\rr_1-\rr_2|)$ refers to the attractive part of fluid-fluid
intermolecular potential. In our study we use a cut-off dispersion-like forces, $\phi_{\rm att}(r)=-4\varepsilon(\sigma/r)^6$ which is truncated at
$r_c=2.5\,\sigma$ where $\sigma$ is the hard-sphere diameter. For the repulsive part of the free-energy we use a highly accurate Rosenfeld-like
hard-sphere free energy functional  $F_{hs}=\int \Phi(\{n_\alpha\})\,\dr$ where the free energy density $\Phi$ is a function of a set of weighted
densities $\{n_\alpha\}$, the explicit expressions of which can be found in Ref.\,\cite{mal13b}.

In addition to the fluid-fluid pair potential the fluid molecules are also subject to an external potential $V(\rr)=V(x,z)$ which arises from the
presence of the pair of parallel capillary walls. The walls are placed in the planes $x=0$ and $x=L$ and, while of finite height $H$, are considered
of infinite length and depth. Assuming that there is uniform distribution of atoms with the density $\rho_w$, the external potential is obtained by
integrating a pair-wise substrate-fluid potential $\phi_{\rm w}$ over the whole domain of the walls. In our study we use a long-ranged dispersion
like potential $\phi_{\rm w}=-4\epsilon_w(\sigma/r)^6$ for $r>\sigma$ for which we find that
 $V(x,z)=\tilde{V}(x,z)+\tilde{V}(L-x,z)$ where $\tilde{V}(x,z)=\alpha_w[\psi(x,z)+\psi(x,H-z)]$, with
 \bb
  \psi(x,z)=\frac{2x^4+x^2z^2+2z^4}{2x^3z^3\sqrt{x^2+z^2}}-\frac{1}{z^3}
 \ee
and $\alpha_w=-\pi/3\varepsilon_w\rho_w\sigma$. Furthermore, the walls are assumed to be impenetrable, so that within the capillary slit an extra hard-wall potential acts for $x<\sigma$ and $x>L-\sigma$. In the limit of $H\to\infty$ and $L\to\infty$ this potential reduces to $V(x,z)=2\alpha_w/x^3$, for $x>\sigma$, pertinent to a single, infinite, planar wall.

\begin{figure}[h]
 \centerline{\includegraphics[width=8cm]{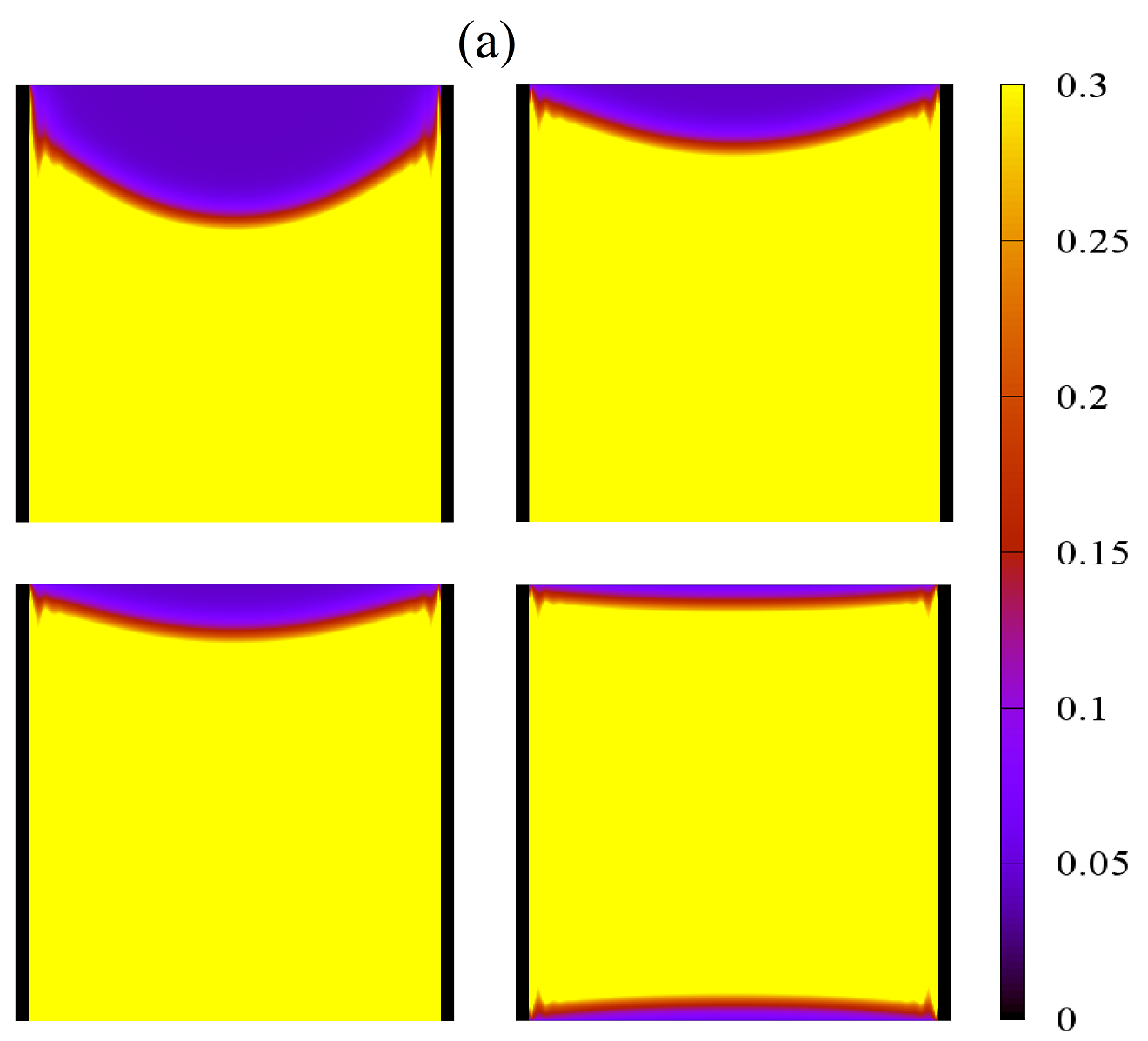}}
 \centerline{\includegraphics[width=8cm]{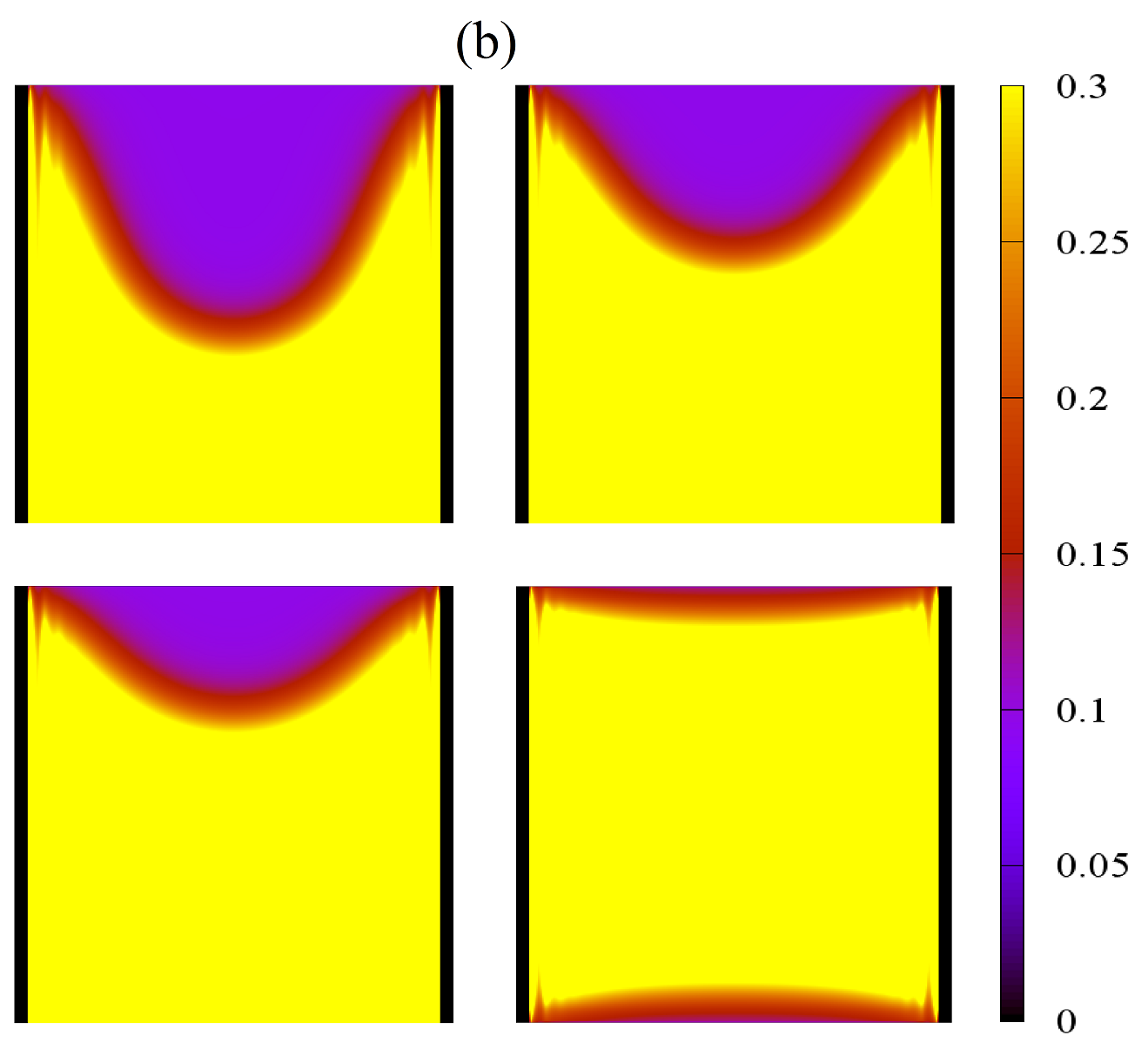}}
 \caption{Two-dimensional
density profiles for the  condensed liquid-like state exactly at capillary condensation for a) partial wetting $\theta\approx 30^\circ$ and b)
complete wetting $\theta=0^\circ$. The profiles shown are for fixed capillary width $L=40\,\sigma$ and for capillary heights (from top left to bottom
right) $H=300\,\sigma, 100\,\sigma, 80\,\sigma$ and $40\,\sigma$. Only the top part of the capillary is shown for $H>40\sigma$.} \label{fig2}
\end{figure}

\begin{figure}[h]
\centerline{\includegraphics[width=10cm]{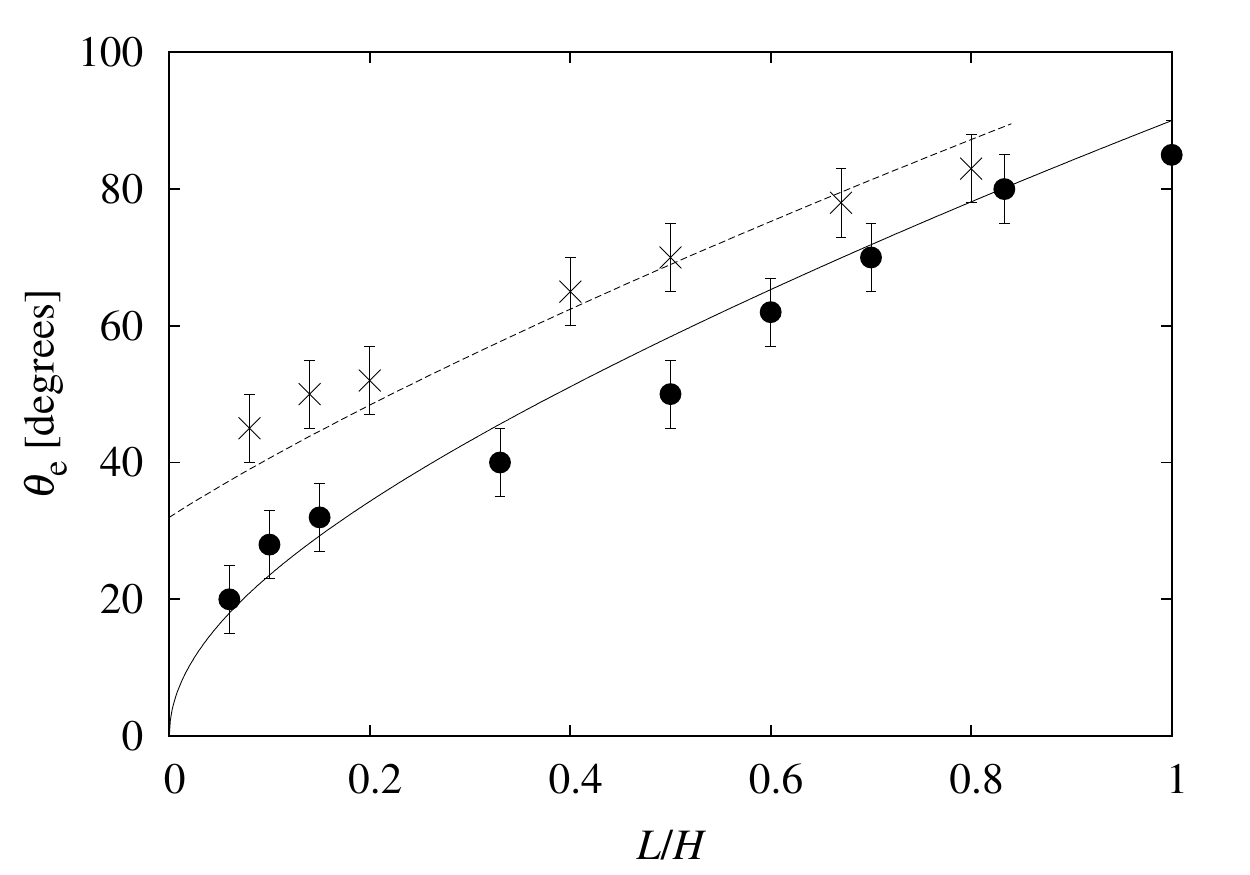}}
 \caption{Values of the edge contact angle for different slit heights obtained from fitting a meniscus shape to the DFT density profiles
 for a slit of width $L=40\,\sigma$.
 Upper symbols refer to $T<T_w$ lower symbols to $T>T_w$. Comparison with the macroscopic prediction (\ref{thetae})
 in each case is shown (dashed and solid lines) and error bars are displayed.} \label{fig3}
\end{figure}

\begin{figure}[h]
\includegraphics[width=9cm]{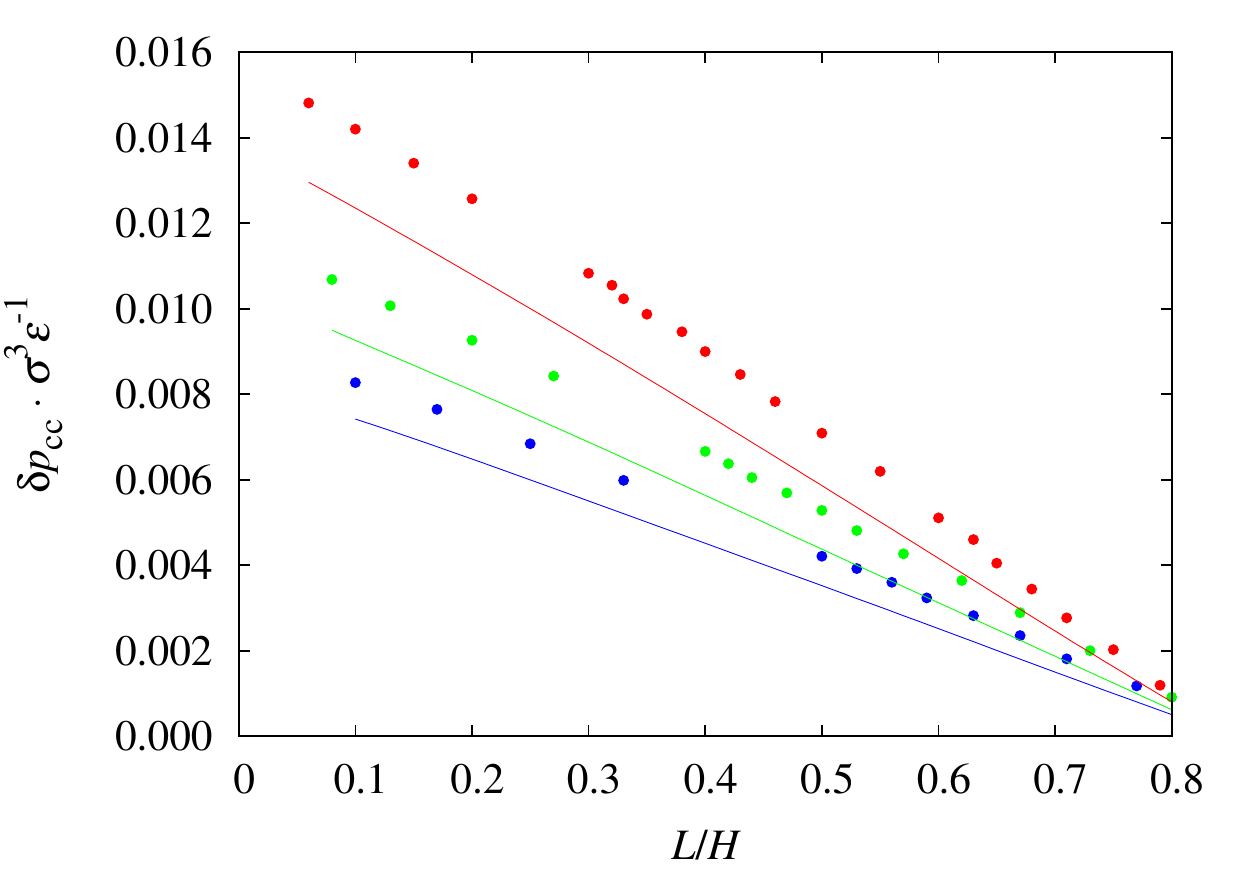}
\caption{Numerical results (dots) for $\delta p_{cc}(L;H)$ for partial wetting ($T<T_w$ and $\theta\approx 30^\circ$) for increasing slit widths,
$L=30\,\sigma$ (red), $L=40\,\sigma$ (green) and $L=50\,\sigma$ (blue), and different heights $H$ compared with the prediction of the modified Kelvin
equation (\ref{mod_kelvin}).}\label{fig4}

\end{figure}

\begin{figure}[h]
\includegraphics[width=8cm]{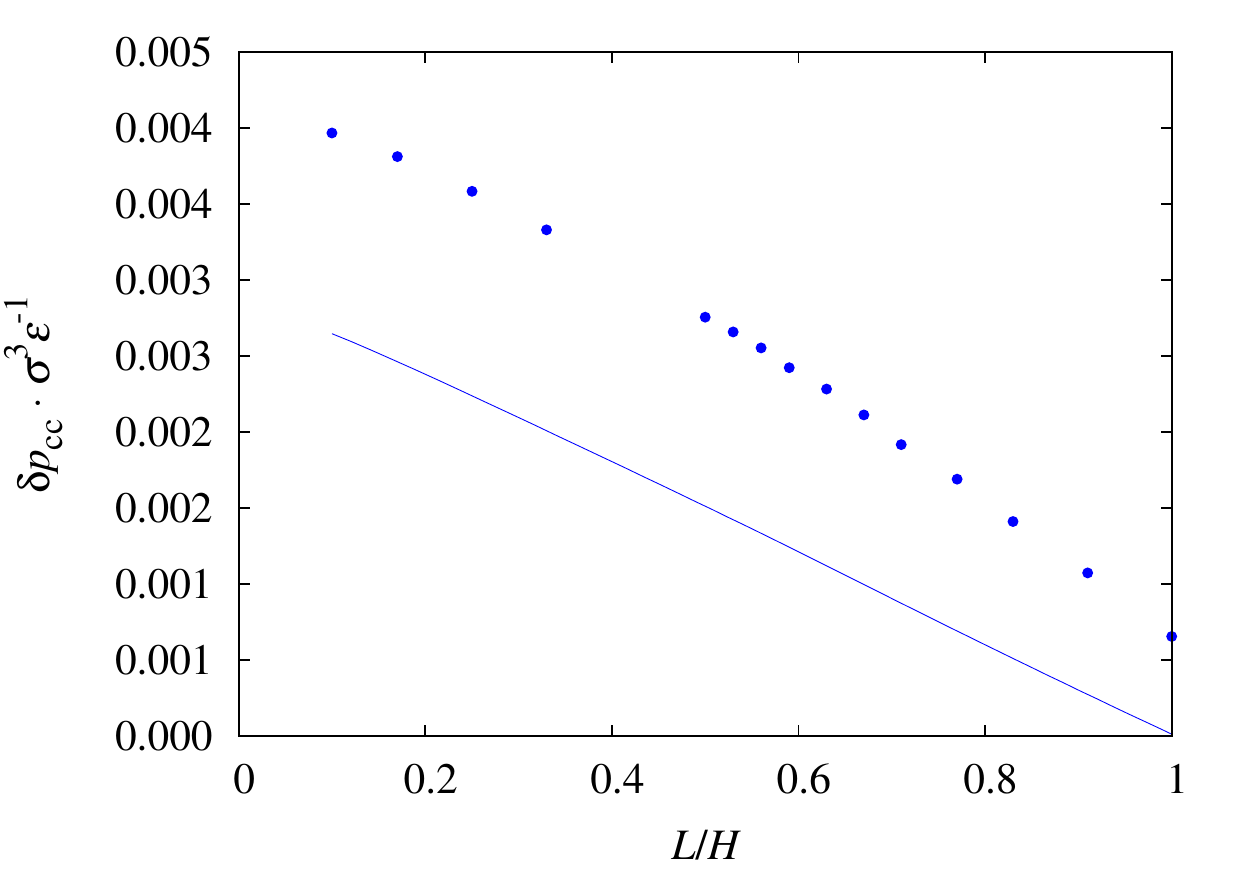}
\caption{Numerical results (dots) for $\delta p_{cc}(L;H)$ for complete wetting ($T>T_w$ and $\theta=0^\circ$) for $L=50\sigma$  and different
heights $H$ compared with the prediction of the modified Kelvin equation (\ref{mod_kelvin}).} \label{fig5}
\end{figure}

This microscopic model has been shown to capture correctly packing effects of fluid atoms in confinement and to obey accurately exact statistical
mechanical sum rules at planar, spherical and wedge-like geometries. The mean-field treatment of attractive forces of course neglects some
fluctuation effects near surface phase transitions but we do not anticipate these to be important for the first-order capillary-condensation
transition considered here. Strictly speaking, beyond mean-field we expect that the first-order capillary condensation transition is rounded, due to
finite-size effects, occurring instead over a pressure range centred on $p_{cc}(L;H)$ of order $e^{-\gamma LH/k_B T}$. However this is already
completely negligible for slits where $L$ and $H$ are more than only a few atomic diameters. Thus we are confident that, away from the direct
vicinity of the bulk critical point, the model functional will accurately predict the location of the (shifted) capillary condensation and also
correctly describe the density profile of the condensed phase from which we can extract, at least semi-quantitatively, an observed edge contact
angle. In the following, we use the parameters $\sigma$ and $\varepsilon$ as our units for length and energy respectively. The Euler-Lagrange
equation $\delta\Omega[\rho]/\delta\rhor=0$ was solved numerically on a rectangular grid with the grid size $0.1\,\sigma$ using Picard's iteration
method. The strength of the wall potential was set to $\varepsilon_w=1.2\,\varepsilon$ which, for a single, planar wall induces a first-order wetting
transition at temperature $k_BT_w/\varepsilon=1.18$. We then considered two representative temperatures below and above $T_w$: Partial wetting at
$k_BT/\varepsilon=1.14$ for which $\theta\approx 30^\circ$ and complete wetting at $k_BT/\varepsilon=1.3$ for which $\theta=0$. This higher
temperature is still far away from the bulk critical point ($k_BT_c/\varepsilon=1.41$). We have considered capillaries with widths ranging from
$L=10\,\sigma$ to to $L=50\,\sigma$ and heights between $H=10\,\sigma$ and $H=500\,\sigma$. In the following, we use the parameters $\sigma$ and
$\varepsilon$ as our units for length and energy respectively.

We begin with predictions for the value of the edge contact angle. While the equilibrium contact angle $\theta$ is well-defined thermodynamically
from Young's equation the edge contact angle is only a mesoscopic concept that can only be extracted approximately. We have done this by first
determining the meniscus shape which we have defined as a surface of constant density midway between gas and liquid values. This shape  is then
fitted by a polynomial from which the contact angle can be determined by simple extrapolation. Fig.~\ref{fig2} shows results obtained for a slit of
width $L=40\,\sigma$ below (a) and above (b) the wetting temperature as the height $H$ is reduced. In each caption the density profile profile
corresponds to that of the condensed liquid-like phase exactly at the pressure of capillary condensation. The latter is itself determined numerically
in standard fashion from the equality of the computed grand-potentials of the coexisting liquid-like and gas-like phases. Qualitatively at least it
is clear that for both temperatures the apparent contact angle increases significantly as the height of the slit is reduced. This is perhaps most
clear above the wetting temperature where even for $H=300\sigma$ the apparent contact angle is significantly greater than zero. For comparison, in
this case Eq.~(\ref{thetae}) predicts that $\theta_e\approx26^\circ$. This is clearer still for the smallest slit size ($H=L=40\sigma$) where we can
see from inspection of the bottom right figure that the apparent contact angle is indeed close to the value $\theta_e=\pi/2$ predicted by
(\ref{thetae}). One can attempt to be more quantitative and define a meniscus shape from the density profiles by for example considering tracing the
contour where the density is mid-way between the bulk liquid and gas values. By extrapolating this meniscus shape to the surface of the walls we can
extract an approximate value for $\theta_e$ for different values of $L/H$. This approach becomes an increasingly precise method for measuring
$\theta_e$ as the slit width is increased keeping the  aspect ratio $L/H$ fixed. Results are shown in Fig.~\ref{fig3} and are in fair agreement with
the predictions of the macroscopic theory even for the relatively small system sizes considered here.

Finally, we have tested the modified Kelvin equation (\ref{mod_kelvin}) against the numerically determined pressure of capillary condensation. Recall
that for infinite slits the original Kelvin equation (\ref{kelvin}) works surprisingly well for partial wetting even down to molecularly sizes. For
complete wetting on the other hand, corrections due to the presence of thick wetting layers means that (\ref{kelvin}) is only accurate for widths $L$
greater than several hundred $\sigma$ \cite{evans90}. The same is true when the height $H$ is finite. In Fig.~\ref{fig4} we show a comparison between
the macroscopic prediction given by Eq.~(\ref{mod_kelvin}) and numerical DFT results for $T<T_w$. In this case the modified Kelvin equation is
reasonably accurate even for the pore widths with $L\approx30\,\sigma$ over a wide range of $H$ values. In contrast, but as expected, the prediction
of the Kelvin equation for $T>T_w$ is only semi-quantitative even for $L\approx50\,\sigma$, see Fig.~\ref{fig5}.

In summary, we have derived a modified form of the Kelvin equation for condensation in open slit and cylindrical pores of finite height $H$  and
shown that the pinned menisci of the condensed liquid-like phase are characterised by a geometry dependent edge contact angle $\theta_e>\theta$. The
increase in the apparent contact angle over equilibrium is largest for the case of complete wetting and is in good agreement with results obtained
using an accurate density functional model. Our calculations can be extended to other experimentally accessible geometries such as  parallel circular
plates. All our predictions can be tested in a laboratory using for example confocal microscopy, which has already been used successfully to study
the meniscus shape (under gravity) in horizontal slits \cite{aarts}.

\begin{acknowledgments}
\noindent This work was funded in by the EPSRC UK grant EP/L020564/1 and the Czech Science Foundation, project 17-25100S.
\end{acknowledgments}


\begin{thebibliography}{99}

\bibitem{thomson} W. Thomson, Phil. Mag. {\bf 42}, 448 (1871).

\bibitem{evans79} R. Evans, Adv. Phys. {\bf 28}, 143 (1979).

\bibitem{tarazona83} P. Tarazona and R. Evans, Mol. Phys. {\bf 48}, 799 (1983).

\bibitem{evans86} R. Evans, U. Marini Bettolo Marconi, and P. Tarazona, J. Chem.
Phys. {\bf 84}, 2376 (1986).

\bibitem{gelb99} L. D. Gelb, K. E. Gubbins, R. Radhakrishnan, and M. Sliwinska-Bartkowiak, Rep. Prog. Phys. {\bf 62}, 1573 (1999).

\bibitem{rascon07} C. Rasc\'on, A. O. Parry, N. B. Wilding, and R. Evans, Phys. Rev. Lett. {\bf 98}, 226101 (2007).

\bibitem{roth11} R. Roth and A. O. Parry, Mol. Phys. {\bf 109}, 1159 (2011).

\bibitem{mal12} A. Malijevsk\'y,  J. Chem. Phys. {\bf 137}, 214704 (2012).

\bibitem{rascon13} C. Rasc\'on, A. O. Parry, R. N\"{u}rnberg, A. Pozzato, M. Tormen, L. Bruschi, and G. Mistura, J. Phys.: Condens. Matter {\bf 25},
192101 (2013).

\bibitem{petr13} P. Yatsyshin, N. Savva, and S. Kalliadasis, Phys. Rev. E {\bf 87}, 020402 (2013).

\bibitem{mal14} A. Malijevsk\'y and A. O. Parry, J. Phys.: Condens. Matter {\bf 26}, 355003 (2014).

\bibitem{rosenfeld} Y. Rosenfeld,  Phys. Rev. Lett. {\bf 63}, 980 (1989).

\bibitem{Luzar} K. Lum and A. Luzar,  Phys. Rev. E {\bf{56}}, R6283 (1997).

\bibitem{Netz} M. Kanduc, A. Schlaich, E. Schneck, and R. R. Netz, Langmuir {\bf{32}}, 8767 (2016).

\bibitem{Chacko} B. Chacko, R. Evans, and A. Archer, J. Chem. Phys. {\bf 141}, 124703 (2017).

\bibitem{mal13} A. Malijevsk\'y and A. O. Parry, Phys. Rev. Lett. {\bf 110}, 166101 (2013).

\bibitem{mal13b} A. Malijevsk\'y, J. Phys.: Condens. Matter {\bf 25},  445006 (2013).

\bibitem{evans90} R. Evans, J. Phys.: Condens. Matter. {\bf 2} 8989, (1990).

\bibitem{aarts}  A. O. Parry, C. Rasc\'on, E. A. G. Jamie, and D. G. A. L.  Aarts, Phys. Rev. Lett.  {\bf 108}, 246101  (2012).


\end{thebibliography}
\end{document}